\documentclass[runningheads]{llncs}
\usepackage[T1]{fontenc}
\usepackage{fullpage}
\usepackage{graphicx}
\usepackage{amsmath}
\usepackage{algorithm}
\usepackage{algorithmic}
\usepackage{enumitem} 
\usepackage{xcolor}
\usepackage{comment}
\newtheorem{definition6}{Definition}
\newtheorem{definition1}{Lemma}
\newtheorem{definition2}{Proposition}
\newtheorem{definition3}{Theorem}
\newtheorem{definition4}{Corollary}

\definecolor{darkgreen}{rgb}{0.0,0.5,0.0}
\definecolor{brown}{rgb}{0.65,0.16,0.16}
\usepackage[utf8]{inputenc}
\usepackage{graphicx} % Required for inserting images
\usepackage{subcaption} % Required for side-by-side subfigures

\begin{document}
	
	\title{Terminal Steiner tree problem : Complexity and Algorithms }
	\titlerunning{Terminal Steiner tree problem : Complexity and Algorithms }
	
	\author{Jyothish S\inst{1} \and
		N Sadagopan\inst{1} 
	}
	\authorrunning{Jyothish S et al.}
	
	\institute{Indian Institute of Information Technology, Design and Manufacturing,
		Kancheepuram, Chennai, India \\
		\email{\{cs24d0012,sadagopan\}@iiitdm.ac.in}}

	\maketitle
	
	\begin{abstract}
		Given a connected graph $G$ and a terminal set $R \subseteq V(G)$, the Steiner tree problem (ST) asks for a tree that spans all of $R$ with at most $r$ vertices from $V(G)\backslash R$, for some integer $r\geq 0$. It is known from (Garey et al.,1977 \cite{garey}) that ST is NP-complete. A Steiner tree in which all terminal vertices are constrained to be leaves is called a terminal Steiner tree. Our study addresses the existence of a terminal Steiner tree, its complexity across various graph classes, black-box applications of the ST, and a fixed-parameter tractable (FPT) algorithm with respect to the number of terminals.

	\end{abstract}
	
	\section{Introduction}
	
	The Steiner tree problem is one of the most important combinatorial optimization problems, which can be used as a model in many fields, such as global routing, network routing, VLSI design, optical and wireless communication systems, transportation and distribution networks, and phylogenetic tree reconstruction. It is one among 21 NP-complete problems listed by Karp \cite{Karp1972}. The Steiner tree problem is a common term for a class of combinatorial optimization problems defined in various settings. Generally speaking, this problem requires an optimum interconnect for a given set of points under a predefined objective function.
	
	Given a connected edge-weighted graph \(G\) and a subset of vertices \(R \subseteq V(G)\) called terminals, the objective is to find a tree \(T\) that spans all vertices in \(R\) while minimizing the total weight of the edges in \(T\). In the unweighted setting, the goal becomes minimizing either the number of edges in the Steiner tree or the number of additional vertices \(Q \subseteq V(G) \setminus R\) (called Steiner vertices) used to connect the terminals. ST generalizes two well-known problems: the Minimum Spanning Tree when \(R = V(G)\), and the Shortest Path problem when \(|R| = 2\). In this paper, we shall work with the unweighted version of the problem. All graphs in this paper are assumed to be simple unweighted connected graphs unless explicitly mentioned otherwise. 
	
	Since ST is a hard problem for general graphs, it has been studied in different graph classes as well as for approximation results. ST remains NP- complete even on graph classes like Planar \cite{GareyJohn}, Bipartite \cite{Karp1972}, Split \cite{white}, and Chordal bipartite graphs \cite{MULLER1987257}. While the problem is solvable in polynomial time for certain graph classes, such as Series Parallel \cite{Seriespar}, Strongly Chordal \cite{white}, Permutation \cite{COLBOURN1990179} and Circle graph \cite{DEFIGUEIREDO2022184}.

	The problem of identifying an optimum tree in which all terminal vertices are leaves is referred to as the Terminal Steiner Tree problem(TST). Even though every connected graph admits a Steiner tree, the existence of a terminal Steiner tree is not assured.  It depends on the choice of the terminal set.
	In the Euclidean setting, a terminal Steiner tree is guaranteed to exist for any terminal set $R$. This follows from the fact that one can always construct a star configuration with the terminal vertices as leaves, introducing additional interior points when collinearity occurs. An analogous construction ensures the existence of a rectilinear terminal Steiner tree.
	
	TST on graphs is NP-complete \cite{articleTST}. If $\rho$ is the best-known approximation ratio for the graph Steiner tree problem, then there is a polynomial time approximation algorithm for Terminal Steiner Tree with performance ratio $\rho+2$ \cite{articleTST}. Later, it improved to $2\rho$ \cite{DRAKE200415}. Current best approximation algorithm for Terminal Steiner tree have performance ratio $2\rho -\frac{\rho \alpha^2-\alpha \rho}{(\alpha+\alpha^2)(\rho-1)+2(\alpha-1)^2}$, $\alpha \geq 2$ \cite{term3}.
	
	We establish the necessary and sufficient conditions for the existence of a Terminal Steiner tree for a particular terminal set in an arbitrary graph. 
	We subsequently present a general framework that characterises the graph classes in which the TST is NP-complete. We propose a framework to solve TST from ST for split-like graphs. We further establish that the weighted version of the Terminal Steiner Tree problem is fixed-parameter tractable (FPT) with respect to the number of terminals. 
	\subsection{Preliminaries}
	
	In this paper, we work with connected, simple, unweighted graphs. For a graph $G$, the vertex set is $V(G)$ and the edge set is $E(G) = \{\{u,v\} | u,v \in V(G) $ and $u$ is adjacent to $v \in G$ and $u\neq v\}$. The neighborhood of a vertex $v$ denoted by $N_G(v)=\{u|\{u,v\}\in E(G)\}$. The degree of a vertex $v$ is $deg_G(v)=|N_G(v)|$. For a graph $G$ and $S\subseteq V(G)$, $G[S]$ represents the subgraph of $G$ induced on the vertex set $S$. The diameter of a graph is the farthest distance between any two of its vertices.
	
	A \emph{bipartite} graph $G=A\cup B$ is such that $V(G)$ can be partitioned into two independent sets $A$ and $B$. $K_{m,n}$ is a \emph{complete bipartite} graph with $|A|=m$ and $|B|=n$. A \emph{split} graph $G=K+ I$ is such that $V(G)$ can be partitioned into a clique $K$ and an independent set $I$.  A \emph{bisplit}  graph $G=[A\cup B]+ I$ is such that $V(G)$ can be partitioned into three independent sets $A$,$B$ and $I$, such that $G[A\cup B]$ is a complete bipartite graph.
	
	We define \emph{split-like graph} as a graph $G$ which can be partitioned into two sets $K$ and $I$, where $I$ is an independent set and the induced subgraph $G[K]$ is a complete graph or complete $k$-partite graph, for some $k\geq 1$. For a complete graph or $k=1,2$, $G$ is a split graph, a bipartite graph and a bisplit graph, respectively. For all other values of $k$, the graph is referred to as a $k$-split graph.
	
	A $ H$-free graph is a graph that does not contain $H$ as an induced subgraph. A graph is \emph{chordal} if every cycle of length greater than 3 has a chord. A graph \( G \) is a \emph{chordal bipartite} graph if \( G \) is bipartite and every cycle of length greater than 4 has a chord.
	
	We now formally define the minimum Steiner tree problem.\\
	\noindent\fbox{%
		\parbox{\textwidth}{%
			Optimization problem- ST (G, R):
			\\Input: A Connected Graph $G$, a terminal set $R \subseteq V(G),$
			\\Question: Find a Steiner tree connecting $R$ with the minimum number of Steiner vertices $S$.
		}%
	}
	
	\noindent\fbox{%
		\parbox{\textwidth}{%
			Decision problem- ST(G,R,$k$):
			\\Input: A Connected graph $G$, a terminal set $R \subseteq V(G)$ and an integer $k$
			\\Question: Is there a Steiner tree consisting of at most $k$ Steiner vertices?
		}%
	}

	\section{Terminal Steiner tree}
	Even though every connected graph admits a Steiner tree, the existence of a terminal Steiner tree is not assured.  It depends on the choice of the terminal set. For example, a path is a connected graph in which all subtrees are themselves paths. Hence, if the terminal set has cardinality at least three, no terminal Steiner tree exists for that terminal set.
	Trivial cases in which a terminal Steiner tree always exists are as follows.
	\begin{itemize}
		\item Consider a connected graph $G$ and a terminal set $R$ in which $degree(u)=1$ $\forall u \in R$. In this case, the ST itself is TST.
		\item Consider a connected graph $G$ and a terminal set $R$ such that $|R|=2$, then ST is nothing but the Shortest Path problem, which is also a TST.
		%\item For a complete graph $K_n$ , T-STREE exist for if $|R|< n$
	\end{itemize}
	
	By definition, every terminal Steiner tree is a Steiner tree; however, the converse does not always. Furthermore, given a graph $G$  and a terminal set $R$, the minimal Steiner tree and the minimal terminal Steiner tree (when the latter exists) may differ in their optimum values(i.e. the number of Steiner vertices). Let us consider the complete graph $K_4$ with a terminal set $R$ consisting of any three vertices. In this case, the minimal Steiner tree is the path $P_3$, which requires no additional Steiner vertices, so the optimum value is $k=0$. However, the minimal terminal Steiner tree is the star $K_{1,3}$, which necessarily introduces one Steiner vertex, giving an optimum value of $k=1$. The cardinality of the solution to TST is always greater than or equal to cardinality of the solution to ST.

	\subsection{Existence of Terminal Steiner tree}
	Before determining the optimum terminal Steiner tree, it is necessary to establish whether such a tree exists in the graph $G$ for the specified terminal set $R$.
	\begin{definition1}{\label{l1}}
		For a connected graph $G$ with terminal set $R$, if $\exists u \in R $ such that $N_G(u)\subseteq R$, then either $|R|=2$ or there doesn't exist a Terminal Steiner tree for $R$.
	\end{definition1}
	\begin{proof}
		Let $u \in R $ such that $N_G(u)\subseteq R$. Then for any Steiner tree $T$ for $R$, $|N_G(u)\cap N_T(u)|\geq 1 $. Let $v$ be such a vertex. Again $v$ has the same property of $u$.\\
		Case 1:   $deg_T(v)= 1$; since $deg_T(u)= 1$ (otherwise $T$ is not terminal Steiner tree), T is path $P_2$ .\\
		Case 2: $deg_T(v) \geq 2$; $T$ is no more terminal Steiner tree.
	\end{proof}
	
	Converse is not always true. Consider the path $P_6$ and $R=\{a_2,a_3,a_5\}$. We already know the terminal Steiner tree doesn't exist in a path if $|R|>2$. But none of the $u\in R$ satisfies $N_G(u)\subseteq R$.
	\begin{definition3}\label{Texists}
		Let $G$ be a connected graph with a terminal set $R$. Then the terminal Steiner tree exists for $R$ iff 
		\begin{enumerate}[label=\alph*)]
			\item every terminal vertex has an adjacent vertex in $V\backslash R$.
			\item $G[V \backslash R]$ is connected.
		\end{enumerate}
	\end{definition3}
	\begin{proof}
		Suppose $T$ is a terminal Steiner tree for $R$. Then by Lemma \ref*{l1},(1) is satisfied. Now removing leaves from $T$ doesn't affect the connectivity. i.e. $G[T \backslash R]$ is connected, implies $G[V \backslash R]$ is connected.\\
		Conversely, suppose (a) and (b) are satisfied. Since $G[V \backslash R]$ is connected, find a spanning tree $T'$ for $G[V \backslash R]$. Each terminal vertex is made adjacent to the corresponding vertices in $T'$. Remove all leaves in the obtained tree that are not terminal vertices. The resultant tree $T$ is a terminal Steiner tree for $R$.
		
	\end{proof}
	
	Since the construction process described in this theorem provides a straightforward polynomial-time algorithm, we may conclude that verifying the existence of a terminal Steiner tree in a connected graph is computationally tractable.
	\begin{definition2}
		The problem of determining whether a terminal Steiner tree exists in a connected graph is solvable in polynomial time.
	\end{definition2}
	
	Depending on the spanning tree $T'$ for $G[V \backslash R]$, there exist different terminal Steiner trees with different Steiner vertices $S$. The terminal Steiner tree problem is a minimization problem for which the number of Steiner vertices(interior vertices) has to be minimized. The decision version is defined as
	\begin{definition6}
		TST($G,R,k$)\\
		Input: A connected graph $G$ with a terminal set $R$ and a non negative integer $k$.\\
		Question: Is there any terminal Steiner tree connecting $R$ with at most $k$ Steiner vertices?
	\end{definition6}
	\subsection{Hardness results}
	
	The metric version of TST is known to be NP-complete \cite{articleTST}. Our approach is to investigate the problem within diverse graph classes. Rather than establishing individual proofs for each class, we introduce a unified framework designed to operate effectively across a broad spectrum of graph classes.
	\begin{definition3}\label{bb1}
		Let $\Pi$ be a graph class. Let $\Pi$ be closed under the operation of adding pendant vertices. Then if ST is NP-complete on $\Pi$, then TST is NP-complete on $\Pi$.
	\end{definition3}
	\begin{proof}
		\textbf{TST is in NP}: Given a certificate $T$, we show that there exists a deterministic polynomial-time algorithm for verifying the validity of the certificate $T$. Note that the standard Breadth First Search algorithm can be used to check whether $T$ is connected and acyclic. It is easy to check whether $|S| \leq k$. Also check $deg_T(v)=1$ $\forall v \in R$ Therefore, the certificate verification can be done in $O(|V(G)| + |E(G)|)$. Thus, we conclude that the Terminal Steiner tree problem is in NP.\\
		\textbf{TST is NP- Hard}: An instance of $ST(G,R,k)$ is reduced to an instance of $TST(G',R',k')$.\\
		\textbf{Construction:} Here $G$ is in $\Pi$ . For each $u\in R$, create a copy $u'$ and make $u'$ adjacent to $u$. Let $R'=\{u': u\in R\}$. Then $G'$ is the new graph with $V(G')=V(G)\cup R'$ and $E(G')=E(G)\cup \{\{u,u'\}:u\in R, u' \in R'\}$.\\ 
		\textbf{$G'$ is in $\Pi$:} $G$ is in $\Pi$. Since $\Pi$ is closed under the operation of adding pendant vertices, $G'$ also belongs to $\Pi$.\\
		\textbf{Claim:} $ST(G,R,k)$ if and only  if $TST(G',R',k'=k+|R|)$.\\
		\textbf{Proof:} Necessity: If there exist a Steiner set $S$ of size at most $k$ in $G$ for $R$, then $S'=S\cup R$ forms a Steiner set of size at most $k+|R|$ for $R'$ ( since $S\cap R=\phi$) and the resultant tree will have $R'$ as leaves.\\
		Sufficiency: Suppose there exists a Steiner set $S'$ of size at most $k+|R|$ in $G'$ for $R'$. Since $R'$ are connected only to $R$, $R \subseteq S'$ .Then the set $S=S'\backslash R$ forms a Steiner set in $G$ for $R$. Also $|S|=|S'\backslash R|=|S'|-|S'\cap R|=|S'|-|R|\leq k+|R|-|R|=k$.
	\end{proof}
	
	Graph classes closed under adding pendant vertices, for which ST is known to be NP-complete, include bipartite \cite{Karp1972}, planar \cite{GareyJohn}, planar bipartite \cite{white}, chordal \cite{white}, and chordal bipartite graphs \cite{MULLER1987257}. By the above theorem,
	\begin{definition4}
		TST is NP-complete on the graph classes- bipartite, planar, planar bipartite, chordal and chordal bipartite.
	\end{definition4}
	
	It should be noted that graph classes like $k$-regular graphs, split graphs and bisplit graphs violate this property, since the addition of pendant vertices takes them outside the class.\\
	The immediate consequences of this construction are described below.
	\begin{definition4}
		Suppose the ST is NP-complete on the graph class of diameter  $d$. Then the TST is NP-complete on the graph class of diameter $d+2$.
	\end{definition4}
	\begin{proof}
		Suppose $G$ is a graph of diameter $d$. Proceed using the same construction as in Theorem \ref{bb1}. Let $P(u,v)$ be a path in $G$ realising the diameter $d$. Adding pendant vertices to $u$ and $v$ increases the path length to $d+2$.\\
		If $u,v\in V(G)$, then $d_{G'}(u,v)=d_{G}(u,v)$\\
		If $u\in V(G)$ and $v'\in R'$ such that $N(v')=\{v\}$, then $d_{G'}(u,v')=d_{G}(u,v)+1$\\
		If $u'\in R'$ such that $N(u')=\{u\}$ and $v'\in R'$ such that $N(v')=\{v\}$, then $d_{G'}(u',v')=d_{G}(u,v)+2$\\
		Therefore \[
		\operatorname{diam}(G') 
		= \max \bigl\{ \operatorname{diam}(G), \operatorname{diam}(G)+1, \operatorname{diam}(G)+2 \bigr\} 
		= \operatorname{diam}(G)+2
		\]
	\end{proof}
	\begin{definition4}
		Assume that ST is NP-complete on the graph class in which $K_{1,r}$ is forbidden as an induced subgraph. Then the TST is NP-complete on the graph class in which $K_{1,r+1}$ is forbidden as an induced subgraph.
	\end{definition4}
	\begin{proof}
		Suppose $G$ is a $K_{1,r}$- free graph. Proceed using the same construction as in the theorem. Assume that the reduced graph $G'$ contains an induced subgraph $H$ which is a $K_{1,r+1}$. If $V(H)\subseteq V(G)$, then it contradicts that $G$ is a $K_{1,r}$- free. So the only possibility is that $H$ is a star with the centre vertex in $G$ and among $r+1$ leaves, at least one vertex is from $R'$. Note that, in the construction, two pendant vertices are never added to the same vertex of $G$. Thus, $H$ is a star whose centre lies in $G$, with exactly one leaf vertex $v\in R'$. Then $H-v$ is a $K_{1,r}$ which is an induced graph in $G$, contradicts that $G$ is $K_{1,r}$- free.
	\end{proof}
	ST is NP-complete in $K_{1,5}$-free split graphs \cite{RENJITH2020246}. An immediate consequence of the above result is that TST is NP-complete in $K_{1,6}$-free split graphs.\\
	We explicitly prove the hardness results for split and bisplit.\\
	
	%Consider the special case of Steiner tree problem in split graph $G=K+I$ such that $R\subseteq I$. Consequently, every Steiner tree in this setting takes the form of a caterpillar, with its backbone entirely contained within $K$. Thus any Steiner tree is a terminal Steiner tree. Hence , in this special case Terminal Steiner tree problem is same as Steiner tree problem.\\
	\noindent\fbox{%
		\parbox{\textwidth}{%
			$EXACT$-$l$-$COVER(X,C)$\\
			Instance: A collection C of 3-element subsets of a set $X = \{x_1,x_2,...,x_{3q}\}$. \\ 
			Question: Is there a sub-collection $C' \subseteq C$ such that for every $x_i \in X$,$x_i$  belongs to exactly one member of $C'$?
		}%
	}
	For $l\geq 3$, often abbreviated as $XlC$, is known to be NP-complete \cite{GareyJohn}.
	\begin{definition2}{\label{p2}}
		TST on a split graph is NP-complete. 
	\end{definition2}
	\begin{proof}
		\textbf{The Terminal Steiner tree problem is NP-Hard:} An instance of $X3C(X,C)$ is reduced to an instance of $TST(G,R,k)$ as follows: $G=K+I$ such that
		$K=\{v_i|c_i\in C\},I=\{u_j|x_j\in X\}$ . $E(G)=\{\{v_i,u_j\} | v_i\in K ,u_j\in I \text{ and }  x_j\in c_i\}\cup \{\{v_i,v_j\}:v_i,v_j \in K, i \neq j \}$ and $R=I$.  \\
		\textbf{$G$ is a connected split graph }with clique $K$.\\
		\begin{figure}[h]
			\centering
			\includegraphics[width=0.7\linewidth]{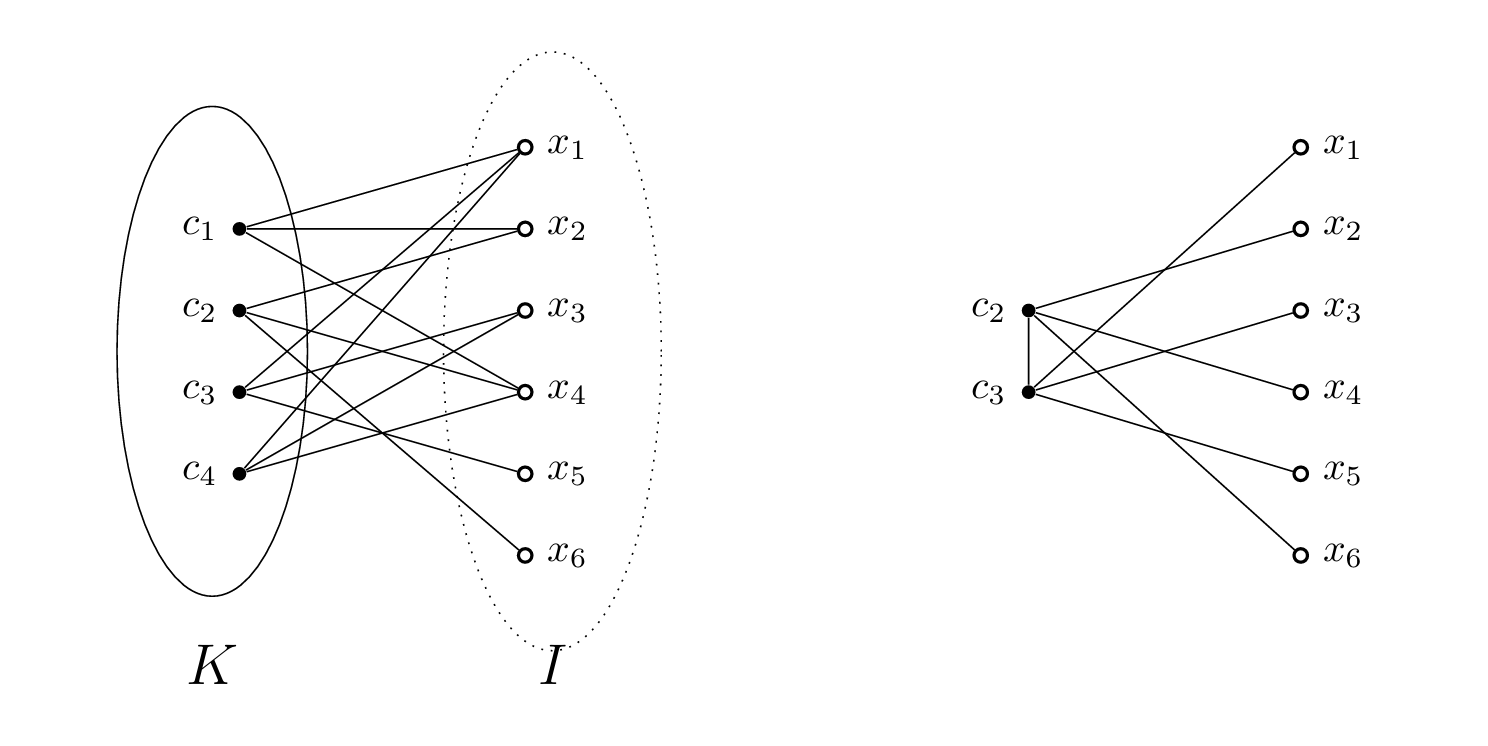}
			\caption{Reduction: An instance of $X3C$ to $TST$ on a split graph and the corresponding minimum terminal Steiner tree with $R=I$.}
			\label{fig:screenshot-2026-02-14-111754}
		\end{figure}
		\textbf{Claim}:$X3C(X,C)$ if and only if $TST(G,R=I\cup I', k=\frac{|X|}{3})$\\
		\textbf{Proof:}\textit{Necessity}: If there exists $C'\subseteq C$,$|C'|=\frac{|X|}{3}$ which covers all the elements of $X$, then the set of vertices $S=\{v_i \in K|c_i \in C \} $ forms a terminal Steiner set in $G$ with $R=I$, where $|S|=\frac{|X|}{3}$.\\
		\textit{Sufficiency}: If there exists a terminal Steiner set $S\subseteq K$ in $G$ on at most $k=\frac{|X|}{3}$ Steiner vertices, then $\forall v\in S, d^{I}(v)=3, |S|=\frac{|X|}{3} \text{ and } N^{I}(S)=|X|$, which implies that there does not exist $u,v \in S$ such that 	$N^{I}(u)\cap N^{I}(v) \neq \emptyset$. Therefore the set $C'=\{c_i\in C|v_i \in S \}$ forms an exact-3-cover of $X$.
		
	\end{proof}
	
	\begin{definition2}{\label{p3}}
		TST on a bisplit graph is NP-complete. 
	\end{definition2}
	\begin{proof}
		\textbf{The Terminal Steiner tree problem is NP-Hard:} An instance of $X3C(X,C)$ is reduced to an instance of $TST(G,R,k)$ as follows: $G=A\cup B+I\cup I'$ such that
		$A=\{v_i|c_i\in C\}$, $B=\{v'_i|c_i\in C\}, I=\{u_j|x_j\in X\}$ and $I'=\{u'_j|x_j\in X\}$. $E(G)=\{\{v_i,u_j\} | v_i\in A, u_j\in I \text{ and }  u_j\in c_i\}\cup \{\{v'_i,u'_j\} | v'_i\in B ,u'_j\in I' \text{ and }  u'_j\in c_i\}\cup \{\{v_i,v'_j\}:v_i \in A, v'_j \in B\}$ and $R=I\cup I'$.  \\
		\textbf{$G$ is a connected bisplit graph} with biclique $A\cup B$.\\
		\textbf{Claim}:$X3C(X,C)$ if and only if $TST(G,R=I,k=\frac{2|X|}{3})$.\\
		\textbf{Proof:}\textit{Necessity}: If there exists $C'\subseteq C$,$|C'|=\frac{|X|}{3}$ which covers all the elements of $X$, then the set of vertices $S=\{v_i \in A|c_i \in C \} \cup \{v_i \in B|c_i \in C \} $ forms a terminal Steiner set in $G$ with $R=I\cup I'$, where $|S|=\frac{2|X|}{3}$.\\
		\textit{Sufficiency}: If there exists a terminal Steiner set $S\subseteq K$ in $G$ on at most $k=\frac{2|X|}{3}$ Steiner vertices, then $\forall v\in S, d^{I\cup I'}(v)=3, |S|=\frac{2|X|}{3} \text{ and } N^{I\cup I'}(S)=|X|$, which implies that there does not exist $u,v \in S$ such that 	$N^{I\cup I'}(u)\cap N^{I\cup I'}(v) \neq \emptyset$. Therefore the set $C'=\{c_i\in C|v_i,v'_i \in S \}$ forms an exact-3-cover of $X$.
		
	\end{proof}
	\subsection{ST as black-box for TST}
	ST on split graph and bisplit graph can be employed as a black-box framework for solving TST. This framework naturally extends to the class of $k$-split graphs.
	\begin{definition2}
		For the class of split graphs in which the ST problem is polynomial-time solvable, the TST problem is also polynomial-time solvable.
	\end{definition2}
	\begin{proof}
		Consider a split graph $G=K+I$. $R \subseteq K$ is trivial. Therefore, our analysis will be concentrated on these two cases.
		\\
		
		Case 1: Suppose $R \subseteq I$.
		Let $T_1$ be a Steiner tree with Steiner set $S_1$ for the terminal set $R$. Without loss of generality, let $u \in R$ such that $deg_{T_1}(u)=2$. Let $ N_{T_1}(u)=\{c_1, c_2\}$. If $c_1, c_2$ are adjacent in $T_1$, it forms a cycle. Therefore, they are non-adjacent. \\
		Case 1(a): Suppose $c_2$ is adjacent to some $w\in I$ (in $T_1$). Then by removing the edge between $c_2$ and $w$, and making the vertices $c_1, c_2$ adjacent in $T_1$. Here, there is no change in optimum. However $degree$ of $u$ in the tree becomes $1$.\\
		Case 1(b): Suppose $c_2$ is not adjacent to any $w \in R$ (in $T_1$). Then $c_2$ can be removed from the optimum, that is $OPT_2 = OPT_1-1$, which is a contradiction. \\
		In this case $OPT_2 = OPT_1$
		\\
		
		Case 2: $R \cap I \neq \phi$ and $R \cap K \neq \phi$.
		This case can be reduced to the case of $R \subseteq I$ by following the method:
		\begin{itemize}
			\item remove edges between two vertices $v_i, v_j \in R\cap K$
			\item remove edges between $v_i \in R\cap K $ and $u_i \in I$
		\end{itemize}
		since these edges will not contribute to the terminal Steiner tree. But the edges between $v_i \in R\cap K $ and $u_i \in I\cap R$ can contribute to the Steiner tree problem. So removing these edges will affect the optimum. For each $v \in R\cap K $, deleting edges from $v$ to other terminal vertices in $I$ may increase the optimum by at most $deg_I(v)$. Thus in this case, $OPT_2 \leq OPT_1 + \sum_{v\in R\cap K}deg_I(v)$.
		
		Each step outlined in the process is computationally feasible and can be executed within polynomial time.
	\end{proof}
	
	\begin{definition4}
		TST on $K_{1,4}$-free split graphs is polynomial-time solvable, whereas TST on $K_{1,5}$-free split graphs is NP-complete.
	\end{definition4}
	\begin{proof}
		From \cite{RENJITH2020246} it is known that ST on $K_{1,4}$-free split graphs is polynomial-time solvable. By the preceding proposition, ST on $K_{1,4}$-free split graphs can be reduced in polynomial time to TST on $K_{1,4}$-free split graphs. Also, the graph constructed in the Proposition \ref{p2} is $K_{1,5}$-free.
	\end{proof}
	\begin{definition2}
		For the class of bisplit graphs in which the ST problem is polynomial-time solvable, the TST problem is also polynomial-time solvable.
	\end{definition2}
	\begin{proof}
		Consider a bisplit graph $G=A\cup B+I$. $R \subseteq A\cup B$ is trivial. Therefore, our analysis will be concentrated on these cases. 
		
		Suppose $R \subseteq I$.
		Let $T_1$ be a Steiner tree with Steiner set $S_1$ for the terminal set $R$. Without loss of generality, let $u \in R$ such that $deg_{T_1}(u)=2$. Let $ N_{T_1}(u)=\{c_1, c_2\}$, which is non adjacent in $T_1$.\\
		\\
		Case 1: $N_{T_1}(u) \subseteq A$ (or $B$)\\
		Case 1(a): $b \in S$ for some $  b \in B$. So $c_1$ and $ c_2$ is connected by $b$ in $T_1$. Then in $T_1$ delete the edge $(c_2,u)$. Now $degree$ of $u$ becomes $1$ and there is no change in optimum.\\
		Case 1(b): $b \notin S, \forall b \in B$. Then we need an additional vertex to make a terminal Steiner tree. So introduce a $b'$ from $B$ to $S$. So $S'=S\cup{b'}$. Now, make each R adjacent to exactly one of $S$ and $b'$ is universal to $S$. Thus $OPT_2=OPT_1+1$.\\
		\\
		Case 2: $N_T(u)\cap A \neq \phi$ and $N_T(u)\cap B \neq \phi$. \\
		Suppose $c_1 \in A$ and $c_2 \in B$. Now delete the edge $(u,c_2)$ and add $(c_1,c_2)$ in $T_1$. Now $degree$ of $u$ becomes $1$ and there is no change in optimum.\\
		
		Suppose $R \cap I \neq \phi$ and $R \cap (A\cup B) \neq \phi$.
		This case can be reduced to the case of $R \subseteq I$ by following the method:
		\begin{itemize}
			\item remove edges between two vertices $v_i, v_j \in R\cap (A\cup B)$
			\item remove edges between $v_i \in R\cap (A\cup B) $ and $u_i \in I$
		\end{itemize}
		since these edges will not contribute to the terminal Steiner tree. But the edges between $v_i \in R\cap (A\cup B) $ and $u_i \in I\cap R$ can contribute to the Steiner tree problem. So removing these edges will affect the optimum. For each $v \in R\cap (A\cup B) $, deleting edges from $v$ to other terminal vertices in $I$ may increase the optimum by at most $deg_I(v)$. Thus in this case, $OPT_2 \leq OPT_1 + \sum_{v\in (A\cup B)\cap R}deg_I(v)+1$.% If $R\cap B= \phi $ then Steiner(R in A) $\subseteq B$. But if Steiner(I)$\subseteq B$. we need extra one from A.
		\\ 
		
		Each step outlined in the process is computationally feasible and can be executed within polynomial time.
	\end{proof}
	\begin{definition4}
		TST on chordal bisplit and chordal bipartite bisplit is polynomial-time solvable.
	\end{definition4}
	\begin{proof}
		From \cite{inbook} it is known that ST on chordal bisplit and chordal bipartite bisplit graphs is polynomial-time solvable. By the preceding proposition, ST on chordal bisplit and chordal bipartite bisplit graphs can be reduced in polynomial time to TST on the respective graphs.
	\end{proof}
	\subsection{Fixed-Parameter Tractability of TST}
	The Dreyfus-Wagner algorithm \cite{Dreyfus1971TheSP}, solves ST problem in
	time \(\mathcal{O}(3^k)\) by using dynamic programming, where $k$ is the number of terminal vertices. The core of the method is a recursive formula that determines the length of a Steiner minimum tree for a terminal set by utilizing the lengths of Steiner minimum trees for all of its proper subsets. We are trying to extend this method for the minimum terminal Steiner tree. Here we are considering the weighted version of the problem. \\
	Given an edge-weighted connected graph $G$ with a terminal set $R$. The procedure begins by computing minimum terminal Steiner trees for all 2-element subsets of $R$. These results are then used to construct minimum terminal Steiner trees  for all 3-element subsets of $R$, and the process continues inductively until a minimum terminal Steiner tree for the full terminal set $R$ is obtained.\\
	Let $S(X \cup \{v\}) $ denote the length of a min TST for $X \cup \{v\}$ and $sp(v, w)$ denote the length of the shortest $v-w$ path.
	\begin{definition1}\label{Ltst}
		Let \( X \subseteq K \), \( X \neq \emptyset \), and \( v \in V \setminus X \). Then
		\[
		S(X \cup \{v\}) 
		= \min_{\substack{w \in V\setminus R \\ X' \subset X }} 
		\Big\{ sp(v, w) 
		+ S(X' \cup \{w\}) 
		+ S\big((X \setminus X') \cup \{w\}\big) \Big\}.
		\]
	\end{definition1}
	\begin{proof}
		Assume that we have an min TST $T$ for $X \cup \{v\}$ where $X \subseteq R$ and $v \in R \setminus X$. Since $v$ is a leaf of $T$, then there is a vertex $w \in V(T)$ such that there is a shortest path $P_{vw}$ that connects $v$ and $w$ in $T$. Hence we have $T = P_{vw} + T'$, where $T'$ is a min TST for $X \cup \{w\}$. Note that $w$ will not be a terminal ($w \notin R$) otherwise $T$ is not a TST. After removing $P_{vw}$ from $T$, $w$ splits the remaining component $T'$ into two edge disjoint subtrees, i.e. for some non trivial subsets $X' \subseteq X$, min TSTs $T_1'$ for $X' \cup \{w\}$ and $T_2'$ for $(X \setminus X') \cup \{w\}$, we have the decomposition $T' = T_1' \cup T_2'$.
		
		Then min TST $T$ for $X \cup \{v\}$ can be computed from min TST $T'$ for $X' \cup \{w\}$ and shortest path $P_{vw}$, $\forall\, w \in V$, $w \notin R$, and $X' \subseteq X$.
		
		\begin{figure}[htbp]
			\centering
			% Left Image
			
			\begin{subfigure}[b]{0.45\textwidth}
				\centering
				\includegraphics[width=\textwidth]{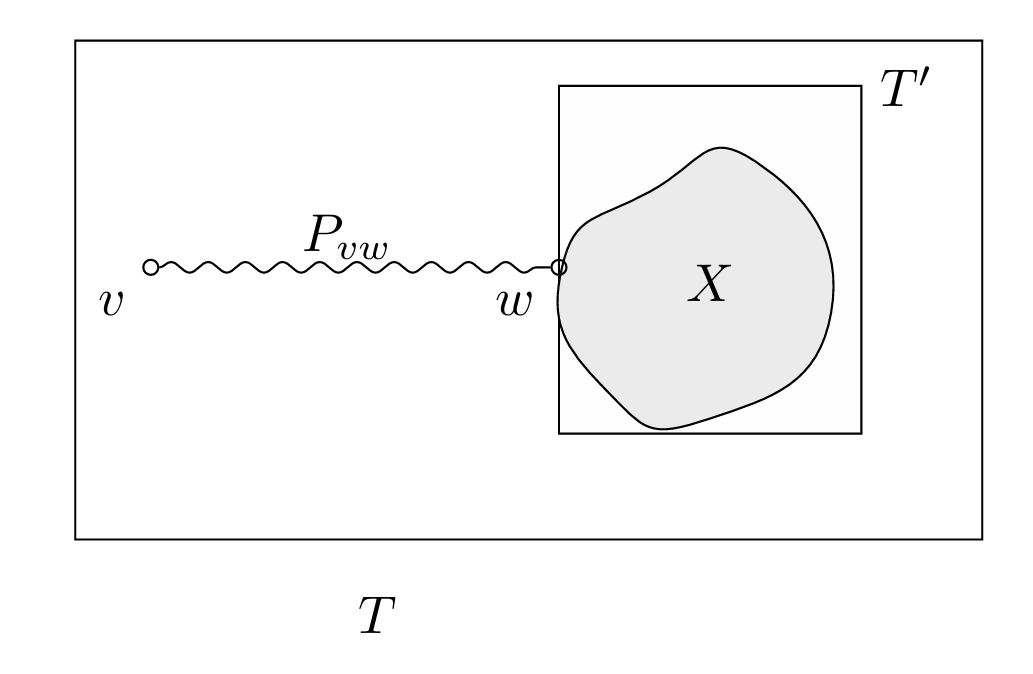}
				\caption{Decomposition of $T = P_{vw} + T'$ }
				\label{fig:image1}
			\end{subfigure}
			\hfill % This adds spacing between the two images
			% Right Image
			\begin{subfigure}[b]{0.45\textwidth}
				\centering
				\includegraphics[width=\textwidth]{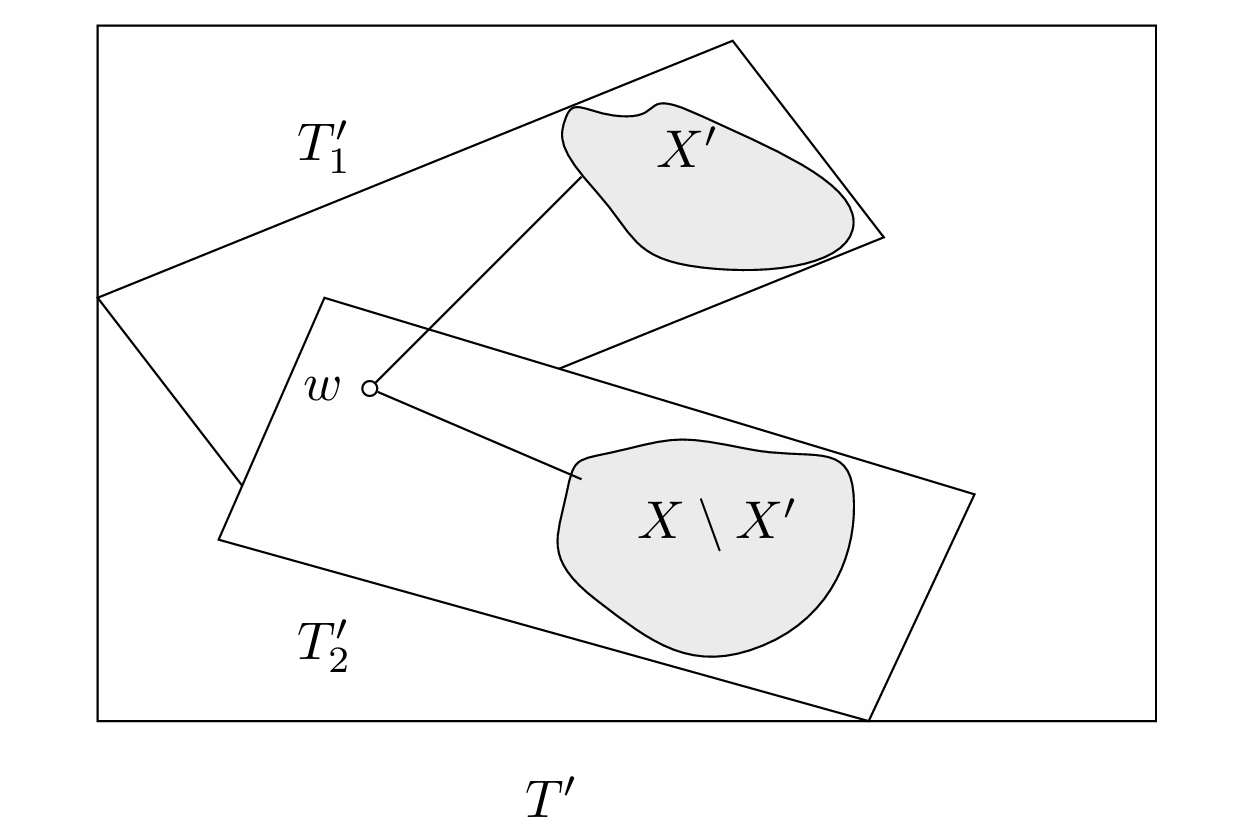}
				\caption{Decomposition $T' = T_1' \cup T_2'$}
				\label{fig:image2}
			\end{subfigure}
			
			\caption{min TST $T$ for $X \cup \{v\}$}
			\label{fig:side_by_side}
		\end{figure}

		Thus, we obtain the following recursion
		\[
		S(X \cup \{v\}) 
		= \min_{\substack{w \in V\setminus R \\ X' \subset X }} 
		\Big\{ sp(v, w) 
		+ S(X' \cup \{w\}) 
		+ S\big((X \setminus X') \cup \{w\}\big) \Big\}.
		\]
	\end{proof}
	Observe that the existence of a TST is guaranteed by the presence of the required non-terminal vertex $w$ in each recursive step.
	\begin{definition3}
		The Dreyfus--Wagner algorithm for TST computes the length of a minimum terminal Steiner tree in \(\mathcal{O}(3^k)\) steps.	
	\end{definition3}
	\begin{proof}
		The correctness of the algorithm follows immediately from Lemma \ref{Ltst}.  The recursion thus allows us to compute all optimal trees $T(X \cup \{v\})$ for
		$v \in  V$ and $X \subseteq R$ of size $|X| = i$ recursively for $i = 1, 2, \cdots, k$.
		Assuming that we have already computed all these trees up to level $i-1$, the minimum
		in the recursion for a given $X \subseteq R$ of size $|X| = i$ can be computed in time $O(2^i)$.
		Hence, in total, the algorithm takes
		$$\mathcal{O}\!\left(\sum_{i=1}^{k} \binom{k}{i} 2^i\right) = \mathcal{O}(3^k).$$
	\end{proof}

	\section{Conclusion}
	We present a characterization that establishes the conditions under which a terminal Steiner tree exists in a connected graph with a specified terminal set, yielding a polynomial-time algorithm for verifying the existence of a terminal Steiner tree. Then we present hardness results for various graph classes.
	We propose a black‑box algorithmic framework for the Terminal Steiner Tree problem, derived from the foundational Steiner tree problem, for split-like graphs. However, for bipartite graphs, the existence of such an algorithm (and bounds) remains unsettled. It is interesting to extend the study of the TST to other graph classes, in order to identify cases where the ST is tractable while TST is intractable, and conversely where TST is tractable but ST is not, which facilitates the development of approximation algorithm for the corresponding intractable problem.
	\bibliography{splitlik} % my bib file references.bib

@article{garey,
	title={A Guide to the Theory of NP-Completeness},
	author={Garey, Michael R and Johnson, David S},
	journal={Computers and intractability},
	pages={641--650},
	year={1979}
}

@Inbook{Karp1972,
	author="Karp, Richard M.",
	editor="Miller, Raymond E.
	and Thatcher, James W.
	and Bohlinger, Jean D.",
	title="Reducibility among Combinatorial Problems",
	bookTitle="Complexity of Computer Computations: Proceedings of a symposium on the Complexity of Computer Computations, held March 20--22, 1972, at the IBM Thomas J. Watson Research Center, Yorktown Heights, New York, and sponsored by the Office of Naval Research, Mathematics Program, IBM World Trade Corporation, and the IBM Research Mathematical Sciences Department",
	year="1972",
	publisher="Springer US",
	address="Boston, MA",
	pages="85--103",
	isbn="978-1-4684-2001-2",
	doi="10.1007/978-1-4684-2001-2_9",
	url="https://doi.org/10.1007/978-1-4684-2001-2_9"
}

@article{RENJITH2020246,
	title = {The Steiner tree in K1,r-free split graphs—A Dichotomy},
	journal = {Discrete Applied Mathematics},
	volume = {280},
	pages = {246-255},
	year = {2020},
	note = {Algorithms and Discrete Applied Mathematics (CALDAM 2016)},
	issn = {0166-218X},
	doi = {https://doi.org/10.1016/j.dam.2018.05.050},
	url = {https://www.sciencedirect.com/science/article/pii/S0166218X18303111},
	author = {P. Renjith and N. Sadagopan}
}

@article{GareyJohn,
	ISSN = {00361399},
	URL = {http://www.jstor.org/stable/2100192},
	author = {M. R. Garey and D. S. Johnson},
	journal = {SIAM Journal on Applied Mathematics},
	number = {4},
	pages = {826--834},
	publisher = {Society for Industrial and Applied Mathematics},
	title = {The Rectilinear Steiner Tree Problem is NP-Complete},
	urldate = {2025-06-23},
	volume = {32},
	year = {1977}
}

@article{white,
	author = {White, Kevin and Farber, Martin and Pulleyblank, William},
	title = {Steiner trees, connected domination and strongly chordal graphs},
	journal = {Networks},
	volume = {15},
	number = {1},
	pages = {109-124},
	doi = {https://doi.org/10.1002/net.3230150109},
	url = {https://onlinelibrary.wiley.com/doi/abs/10.1002/net.3230150109},
	eprint = {https://onlinelibrary.wiley.com/doi/pdf/10.1002/net.3230150109},
	year = {1985}
}

@article{inbook,
	title = {Bisplit graphs — A structural and algorithmic study},
	journal = {Discrete Applied Mathematics},
	volume = {389},
	pages = {243-253},
	year = {2026},
	issn = {0166-218X},
	doi = {https://doi.org/10.1016/j.dam.2026.04.006},
	url = {https://www.sciencedirect.com/science/article/pii/S0166218X2600209X},
	author = {A. Mohanapriya and P. Renjith and N. Sadagopan}
}

@article{DEFIGUEIREDO2022184,
	title = {Revising Johnson’s table for the 21st century},
	journal = {Discrete Applied Mathematics},
	volume = {323},
	pages = {184-200},
	year = {2022},
	note = {LAGOS’19 – X Latin and American Algorithms, Graphs, and Optimization Symposium – Belo Horizonte, Minas Gerais, Brazil},
	issn = {0166-218X},
	doi = {https://doi.org/10.1016/j.dam.2021.05.021},
	url = {https://www.sciencedirect.com/science/article/pii/S0166218X21002109},
	author = {Celina M.H. {de Figueiredo} and Alexsander A. {de Melo} and Diana Sasaki and Ana Silva}
}

@article{MULLER1987257,
	title = {The NP-completeness of steiner tree and dominating set for chordal bipartite graphs},
	journal = {Theoretical Computer Science},
	volume = {53},
	number = {2},
	pages = {257-265},
	year = {1987},
	issn = {0304-3975},
	doi = {https://doi.org/10.1016/0304-3975(87)90067-3},
	url = {https://www.sciencedirect.com/science/article/pii/0304397587900673},
	author = {Haiko Müller and Andreas Brandstädt}
}

@article{Seriespar,
	title = "Steiner trees, partial 2–trees, and minimum IFI networks",
	author = "Wald, \{Joseph A.\} and Colbourn, \{Charles J.\}",
	year = "1983",
	doi = "10.1002/net.3230130202",
	language = "English (US)",
	volume = "13",
	pages = "159--167",
	journal = "Networks",
	issn = "0028-3045",
	publisher = "Wiley-Liss Inc.",
	number = "2",
}

@article{COLBOURN1990179,
	title = {Permutation graphs: Connected domination and Steiner trees},
	journal = {Discrete Mathematics},
	volume = {86},
	number = {1},
	pages = {179-189},
	year = {1990},
	issn = {0012-365X},
	doi = {https://doi.org/10.1016/0012-365X(90)90359-P},
	url = {https://www.sciencedirect.com/science/article/pii/0012365X9090359P},
	author = {Charles J. Colbourn and Lorna K. Stewart}
}

@article{articleTST,
	author = {Lin, Guo-Hui and Xue, Guoliang},
	year = {2002},
	month = {10},
	pages = {103-107},
	title = {On the terminal Steiner tree problem},
	volume = {84},
	journal = {Inf. Process. Lett.},
	doi = {10.1016/S0020-0190(02)00227-2}
}

@InProceedings{term3,
	author="Chen, Yen Hung",
	editor="Murgante, Beniamino
	and Gervasi, Osvaldo
	and Iglesias, Andr{\'e}s
	and Taniar, David
	and Apduhan, Bernady O.",
	title="An Improved Approximation Algorithm for the Terminal Steiner Tree Problem",
	booktitle="Computational Science and Its Applications - ICCSA 2011",
	year="2011",
	publisher="Springer Berlin Heidelberg",
	address="Berlin, Heidelberg",
	pages="141--151",
	isbn="978-3-642-21931-3"
}

@article{DRAKE200415,
	title = {On approximation algorithms for the terminal Steiner tree problem},
	journal = {Information Processing Letters},
	volume = {89},
	number = {1},
	pages = {15-18},
	year = {2004},
	issn = {0020-0190},
	doi = {https://doi.org/10.1016/j.ipl.2003.09.014},
	url = {https://www.sciencedirect.com/science/article/pii/S0020019003004526},
	author = {Doratha E. Drake and Stefan Hougardy}
}

@article{Dreyfus1971TheSP,
	title={The steiner problem in graphs},
	author={Stuart E. Dreyfus and Robert A. Wagner},
	journal={Networks},
	year={1971},
	volume={1},
	pages={195-207},
	url={https://api.semanticscholar.org/CorpusID:27868355}
}
	\bibliographystyle{ieeetr}
	
\end{document}